\documentclass{ws-ijmpe}

\begin{document}

\markboth{D. Elia et al.}{The pixel Fast-OR signal for the ALICE trigger in p-p collisions}

\catchline{}{}{}{}{}

\title{THE PIXEL FAST-OR SIGNAL FOR THE ALICE TRIGGER \\
IN PROTON-PROTON COLLISIONS}



\author{\footnotesize D. ELIA$^{a,}$\footnote{Corresponding author on behalf of the
SPD project in the ALICE Collaboration.} ,
G. AGLIERI RINELLA$^b$,
A. KLUGE$^b$,
M. KRIVDA$^{b,c}$,
M. NICASSIO$^{a,d}$
}

\address{$^a$ Istituto Nazionale di Fisica Nucleare Sez. di Bari, Bari, Italy \\
$^b$ CERN, European Organization for Nuclear Research, Geneva, Switzerland \\
$^c$ Institute of Experimental Physics, Kosice, Slovakia \\
$^d$ Dipartimento IA di Fisica del Politecnico e dell'Universit\`a, Bari, Italy \\
\vspace{2mm}
{\normalsize\tt Domenico.Elia@ba.infn.it}}

\maketitle

\begin{history}
\received{(received date)}
\revised{(revised date)}
\end{history}

\begin{abstract}
The silicon pixel detector of the ALICE experiment at LHC
comprises the two innermost layers in the inner tracking 
system of the apparatus. It contains 1200 readout chips,
each of them corresponding to a 8192 pixel matrix.
The single chip outputs a digital Fast-OR signal which is
active whenever at least one of the pixels in the matrix
records a hit. 
\\
The 1200 Fast-OR signals can be used to
implement a triggering capability: few details on
the pixel trigger system and some of the
possible applications for the event selection in p-p collisions  
are presented.
\end{abstract}

\section{Introduction}

ALICE is a general-purpose heavy-ion experiment designed to study
the physics of strongly interacting matter and the properties of the quark-gluon
plasma in nucleus-nucleus collisions at LHC.
The ALICE apparatus has several features that make it also an
important contributor to proton-proton physics. A complete
reaserch program on p-p collisions is planned: it aims both
to set the baseline for the understanding of the heavy-ion data
and to explore the new energy domain\cite{1,2}.
\\
The two innermost layers of the ALICE inner tracking system, placed
at 3.9 cm ($|\eta|$$<$1.95) and 7.6 cm ($|\eta|$$<$1.5) average
distances from the beam line, are equipped with hybrid pixels and
constitute the Silicon Pixel Detector (SPD)\cite{1,3}. 
The SPD is composed by 120 basic modules (half-staves), 40 in the
inner layer and 80 in the outer layer. 
Two silicon sensor matrices (ladders) on each half-stave are bump-bonded
to 10 readout chips for a total of 1200 chips for the entire SPD
(400 in the inner layer and 800 in the outer layer).
In each chip the prompt discriminator outputs are OR-ed to generate
a Fast-OR pulse whenever one or more pixel cells record a hit: this
feature can be used to implement a unique triggering
capability for ALICE. 
A schematic illustration of the SPD in the ALICE apparatus and some
details of one of the 120 half-staves are shown in Fig. 1.

\begin{figure}[th]
\centerline{\psfig{file=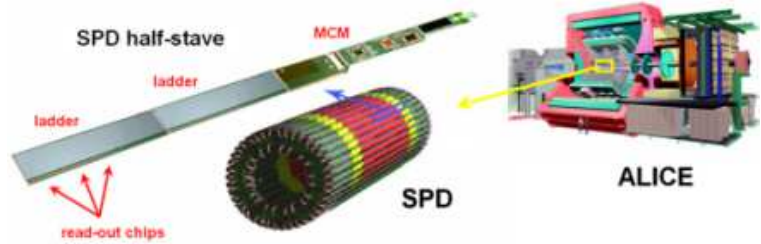,width=10cm}}
\vspace*{2pt}
\caption{The SPD in ALICE and some details of a single
half-stave: sensor ladders, readout chips and Multi Chip Module (MCM).}
\end{figure}
\vspace{-4mm}

In the following sections, after a short description
of the Pixel Trigger system (PIT) architecture, some Fast-OR
based algorithms for triggering in p-p collisions will be presented
and discussed.

\section{Pixel Trigger system architecture}

The 10 Fast-OR bits of each half-stave are continuously transmitted
on the output optical link by the MCM to 
the control room, with a clock frequency of 10 MHz
(bunch crossing frequency in heavy-ion operation\cite{1}).
The PIT has to extract and process all the
1200 Fast-OR signals in order to be able to deliver an input signal for
the Level 0 trigger decision in the ALICE Central Trigger Processor
(CTP). The system targets a maximum latency of 800 ns from the
interaction to the input to the CTP, matching the required 
latency for the Level 0 trigger\cite{4}. 
The architecture of the PIT is schematically shown in Fig. 2. 
The 120 optical fibers from the SPD are connected to an optical splitter 
close to the CTP: data are sent both to the readout electronics
in control room and to the electronics boards of the PIT. 

\begin{figure}[th]
\centerline{\psfig{file=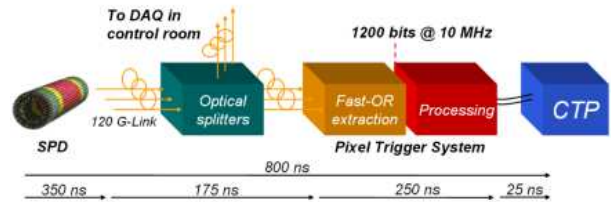,width=8cm}}
\vspace*{2pt}
\caption{The architecture of the Pixel Trigger system.}
\end{figure}
\vspace{-4mm}

The first block of the PIT de-serializes the optical data and extracts 
the 1200 Fast-OR bits from the data flow. The last block implements the processing 
algorithm and generates the input to the CTP: it is constituted by an
electronic board based on a FPGA with a large number of pins and large
logical capacity\cite{5}.

\section{Fast-OR based algorithms}

Studies have been carried out within the ALICE simulation and
reconstruction framework (AliRoot\cite{6}) in order to evaluate the
performance of Fast-OR based algorithms for contributing to event
selection in p-p collisions. In particular triggers for minimum bias events,
high multiplicity events and events with primary tracks in the acceptance
of the High Momentum Particle Identification Detector (HMPID) have been
taken into account. 
The basic event sample used consists of 50000
p-p collisions at $\sqrt{s}=$14 TeV generated by Pythia\cite{7}. 
Fig. 3 shows the corresponding multiplicity
distribution of charged particles, in the pseudorapidity range $|\eta|$$<$1.5.

\vspace{-5mm}
\begin{figure}[th]
\centerline{\psfig{file=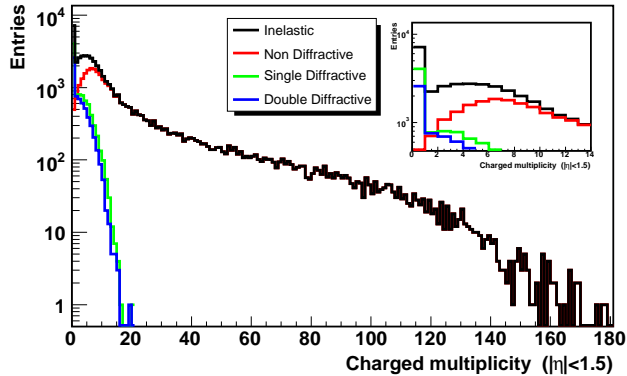,width=9cm}}
\vspace*{2pt}
\caption{Distribution of charged particle multiplicity in $|\eta|$$<$1.5
as generated by Pythia in p-p collisions at 14 TeV for different process types.}
\end{figure}
\vspace{-4mm}

The sample contains single and double diffractive (SD and DD respectively) 
as well as non diffractive events (ND) with the following percentages:
17.8\% SD, 12.7\% DD and 69.5\% ND. All primary charged particles, including 
products of strong and electromagnetic decays, have been taken into account.
Details on the three above mentioned applications of the Fast-OR trigger 
are given in the following sub-sections.

\subsection{Minimum bias trigger}

The minimum bias trigger has to select p-p collisions with the highest efficiency and
the lowest bias, allowing a good rejection of the background (mostly coming from
interaction of the incoming protons with the residual gas in the beam pipe).
Two independent arrays of scintillator counters (VZERO detector\cite{1,8}),
placed at both sides of the interaction point along the beam line, have been designed for
minimum bias trigger and background rejection. Combining the VZERO
trigger signals with the SPD Fast-OR is useful due to the complementarity of the two 
detectors in the geometrical acceptance. 
Different algorithms have been investigated
and the condition of requiring at least one Fast-OR signal in the whole SPD
({\tt GlobalFO}) has shown a very good performance\cite{9}. The following
combinations have been taken into account:

\hspace{15mm}        {\tt MB1 = (GlobalFO .OR.  VZERO-OR)  .AND. notBG} \\
\indent\hspace{15mm} {\tt MB2 = (GlobalFO .AND. VZERO-OR)  .AND. notBG} \\ 
\indent\hspace{15mm} {\tt MB3 = (GlobalFO .AND. VZERO-AND) .AND. notBG} \\
\indent\hspace{15mm} {\tt MB4 = (GlobalFO .OR.  VZERO-AND)  .AND. notBG} \\

\noindent where, based on the arrival time of the particles in one or both
the counters, interactions happening within the two stations
are selected ({\tt VZERO-OR} and {\tt VZERO-AND} conditions) and
those happening outside are rejected ({\tt notBG}).
To check the background rejection capability of such triggers, p-O 
collisions simulated by Hijing\cite{10} have been
also reconstructed and analyzed. As an upper limit estimate of the beam-halo 
contribution to the background, beam-gas collisions occurring
more than 20~m away from the nominal p-p interaction point 
have been taken into account\cite{9}. 
The fractions of p-p and background events
selected by each trigger combination are summarized in Table 1. 

\vspace{-4mm}
\begin{table}[h]
\tbl{Efficiency (in \%) of the different trigger conditions.}
{\begin{tabular}{@{}ccccccc@{}} \toprule
Trigger & Non Diff & Single Diff & Double Diff & All inel & Beam-gas & Beam-halo \\ \colrule
{\tt MB1} &            100.0 & 72.2 & 86.1 & 93.3 & 8.1 & 2.7 \\
{\tt MB2} & \hphantom{0}99.3 & 58.6 & 65.8 & 87.8 & 5.4 & 1.2 \\
{\tt MB3} & \hphantom{0}97.7 & 39.2 & 41.9 & 80.2 & 0.2 & $<10^{-3}$ \\
{\tt MB4} & \hphantom{0}99.7 & 59.5 & 70.9 & 88.9 & 5.6 & 1.5 \\ \botrule
\end{tabular}}
\begin{tabnote}
{\em p-p and background
events are expected to have quite different rates, in parti\-cular
the rate for beam-gas collisions should be one order
of magnitude smaller\cite{2}.}
\end{tabnote}
\end{table}
\vspace{-4mm}

The effect of the trigger selection on various physics distributions
has been studied. As an example, in Fig. 4 the trigger bias on the 
generated track variables $\eta$ and $p_T$ is shown. 

\vspace{-4mm}
\begin{figure}[th]
\centerline{\hspace{5mm}\psfig{file=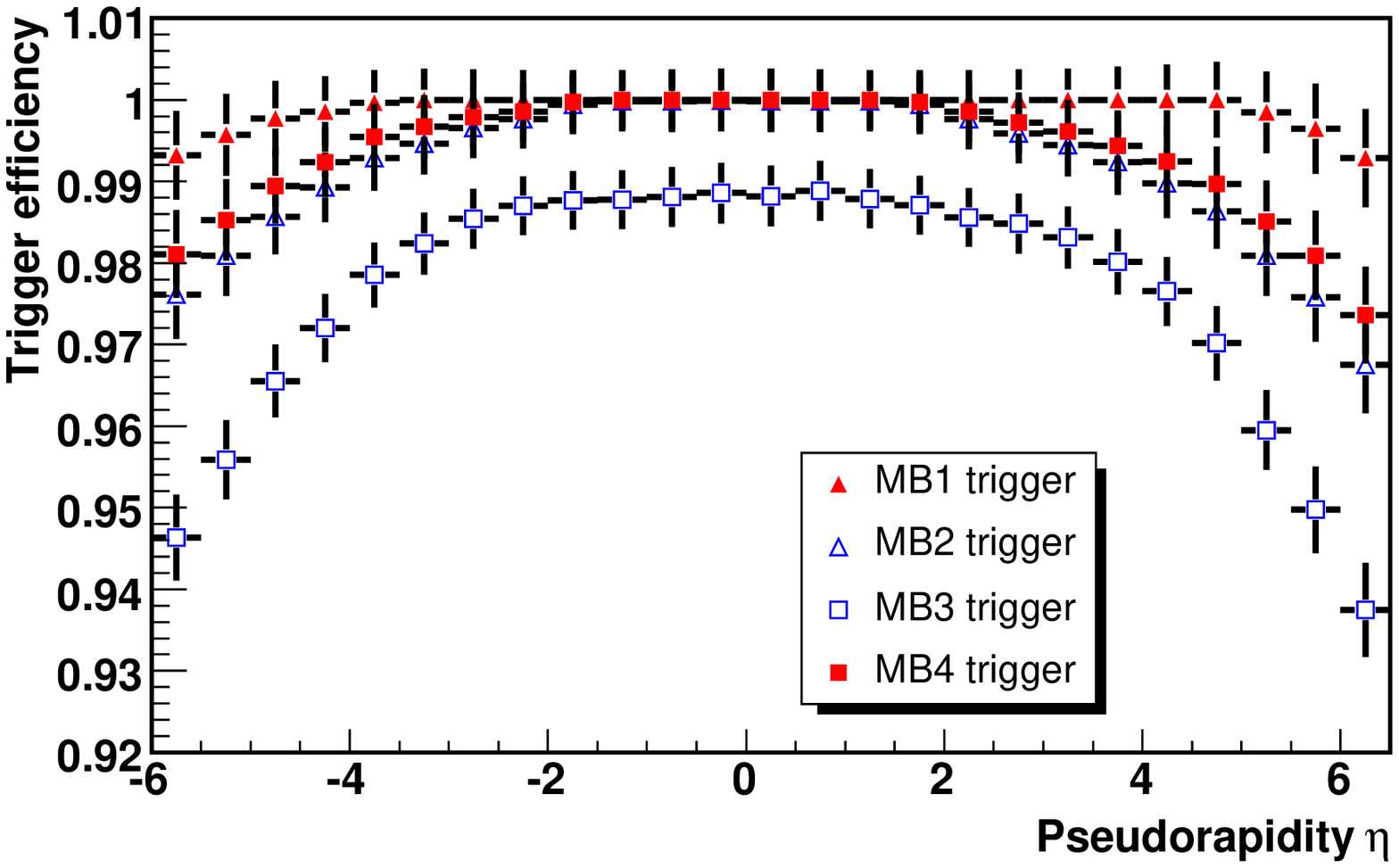,width=6cm}
\psfig{file=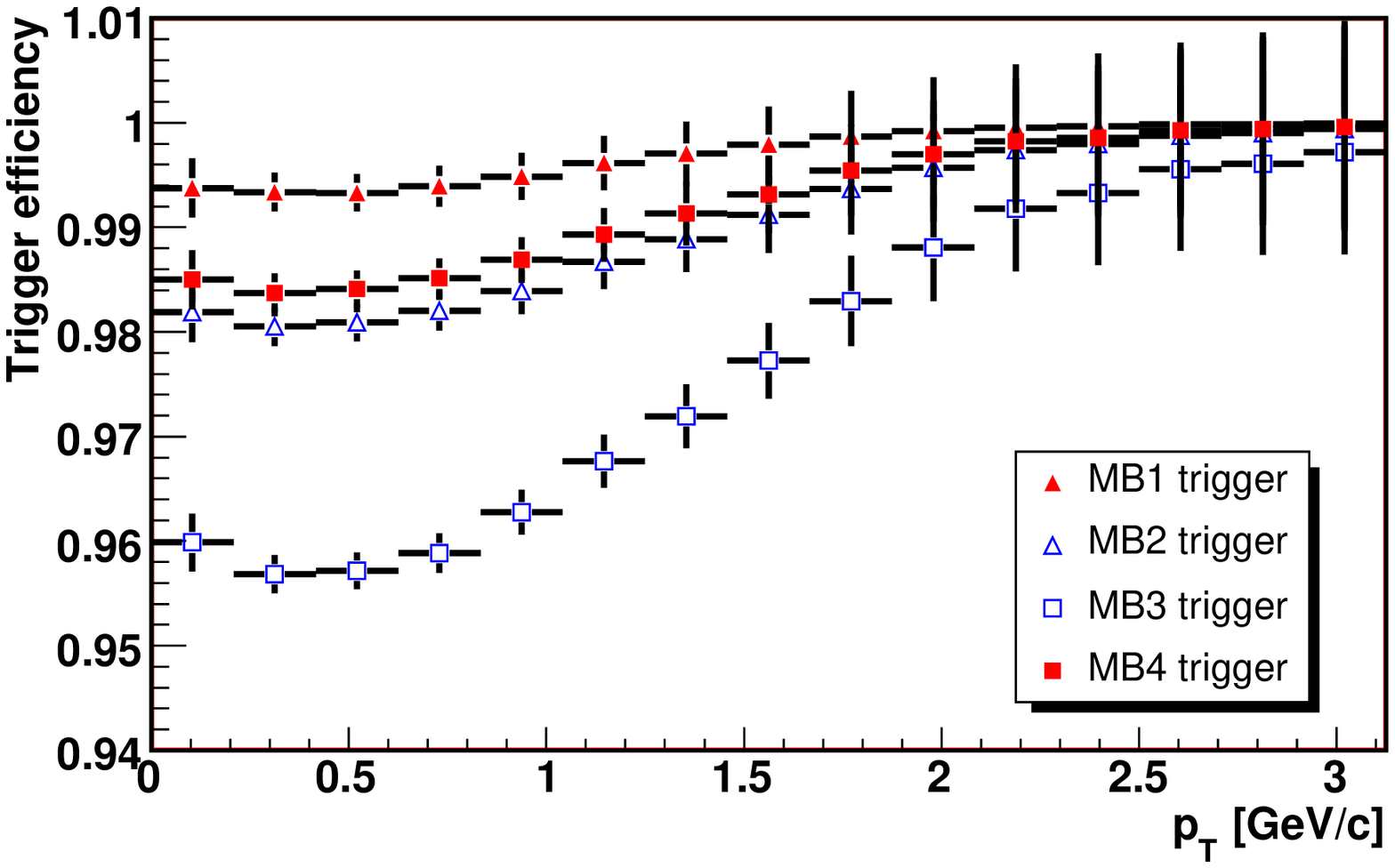,width=6cm}}
\vspace*{2pt}
\caption{Trigger bias on pseudorapidity (left) and
transverse momentum (right) distributions.}
\end{figure}
\vspace{-4mm}

In the central $\eta$ region
covered by the SPD the effect on the pseudorapidity distribution
is negligible for all the triggers but {\tt MB3}: in this last case
the bias, essentially due to the {\tt VZERO-AND} condition,
is less than 2\%. Distortions on the transverse momentum spectrum
are also within few percents.   
In conclusion, triggers {\tt MB1} and {\tt MB4}
are the most efficient with p-p collisions and will be fine for
background rates not far from the current estimates. {\tt MB2}
is still very good as an alternative option preserving bunch 
crossing identification. Finally {\tt MB3} is the most powerful 
in the background rejection and produces a limited bias at event
and track level.

\subsection{High multiplicity trigger}

A Fast-OR based trigger can be used to select high multiplicity events: 
this would allow to study the evolution 
of some physics observables (e.g. $p_T$ spectra, strangeness content) with a 
reasonable statistics.  
The left panel in Fig. 5 shows the correlation between the number of active
Fast-OR signals on the inner SPD layer ({\tt nFOinn}) and the charged multiplicity in the
corresponding pseudorapidity range $|\eta|$$<$2. The panel on the right 
illustrates the effect on the multiplicity spectrum when events are
selected applying lower cuts (30, 60 and 90) in {\tt nFOinn}.

\vspace{-4mm}
\begin{figure}[th]
\centerline{\hspace{5mm}\psfig{file=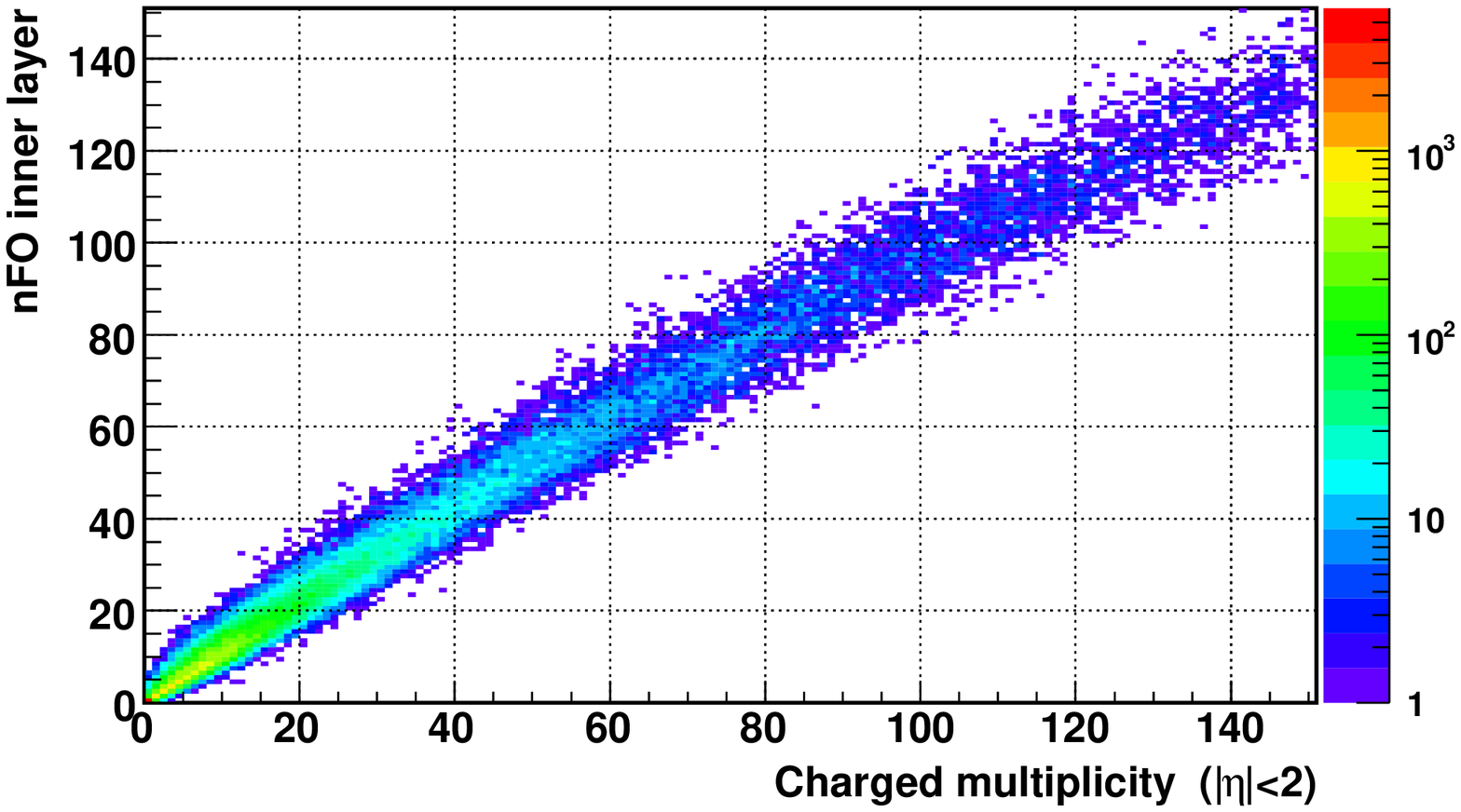,width=6cm}
\psfig{file=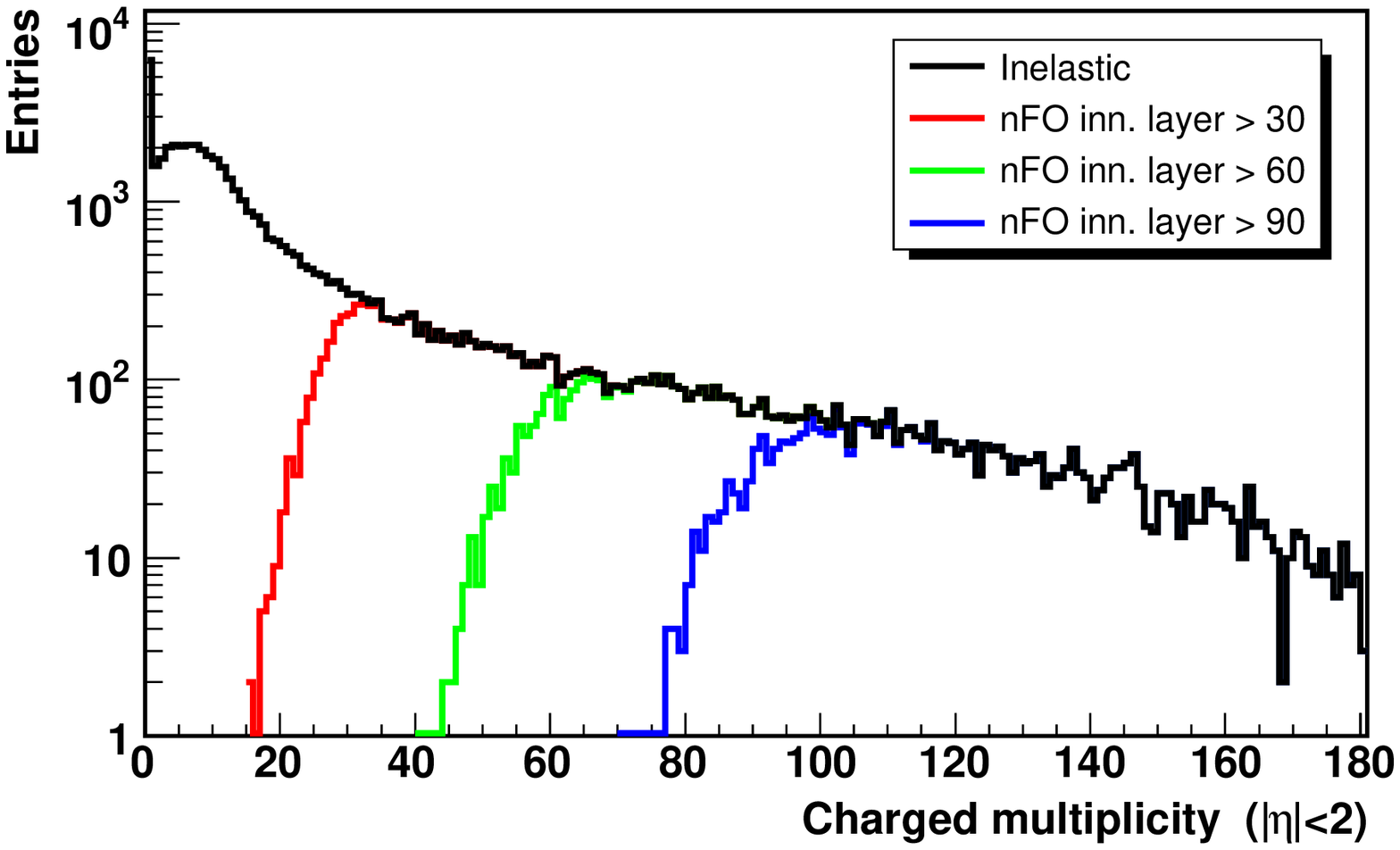,width=6cm}}
\vspace*{2pt}
\caption{Correlation between number of active Fast-OR signals and 
charged multiplicity (left)
and effect of applying lower cuts in {\tt nFOinn} (right).}
\end{figure}

\vspace{-4mm}
High multiplicity triggers have to include VZERO conditions to allow 
background rejection, so they can be defined as:
\\

\hspace{15mm} {\tt HM = (nFOinn $>$ nFOmin) .AND. VZERO-AND .AND. notBG} \\

\noindent ensuring a low residual background contamination ($<$10$^{-4}$). 
How large to set the threshold depends on various
background sources and is currently under study.

\subsection{Trigger for tracks in the HMPID}

Due to its small  acceptance coverage the HMPID is 
traversed by charged primary tracks only in about 10\% of the 
Pythia minimum bias p-p events\cite{1}. 
Increasing the fraction of events with tracks throughout the detector 
would be certainly very useful for some calibration issues
and possibly for dedicated physics studies.
\\
Primary tracks reaching the HMPID are expected to traverse the SPD chips in
defined regions. This is confirmed by the
simulation: Fig. 6 illustrates the location of 
the active Fast-OR signals produced by those tracks on the two SPD
layers. 

\vspace{-4mm}
\begin{figure}[th]
\centerline{\psfig{file=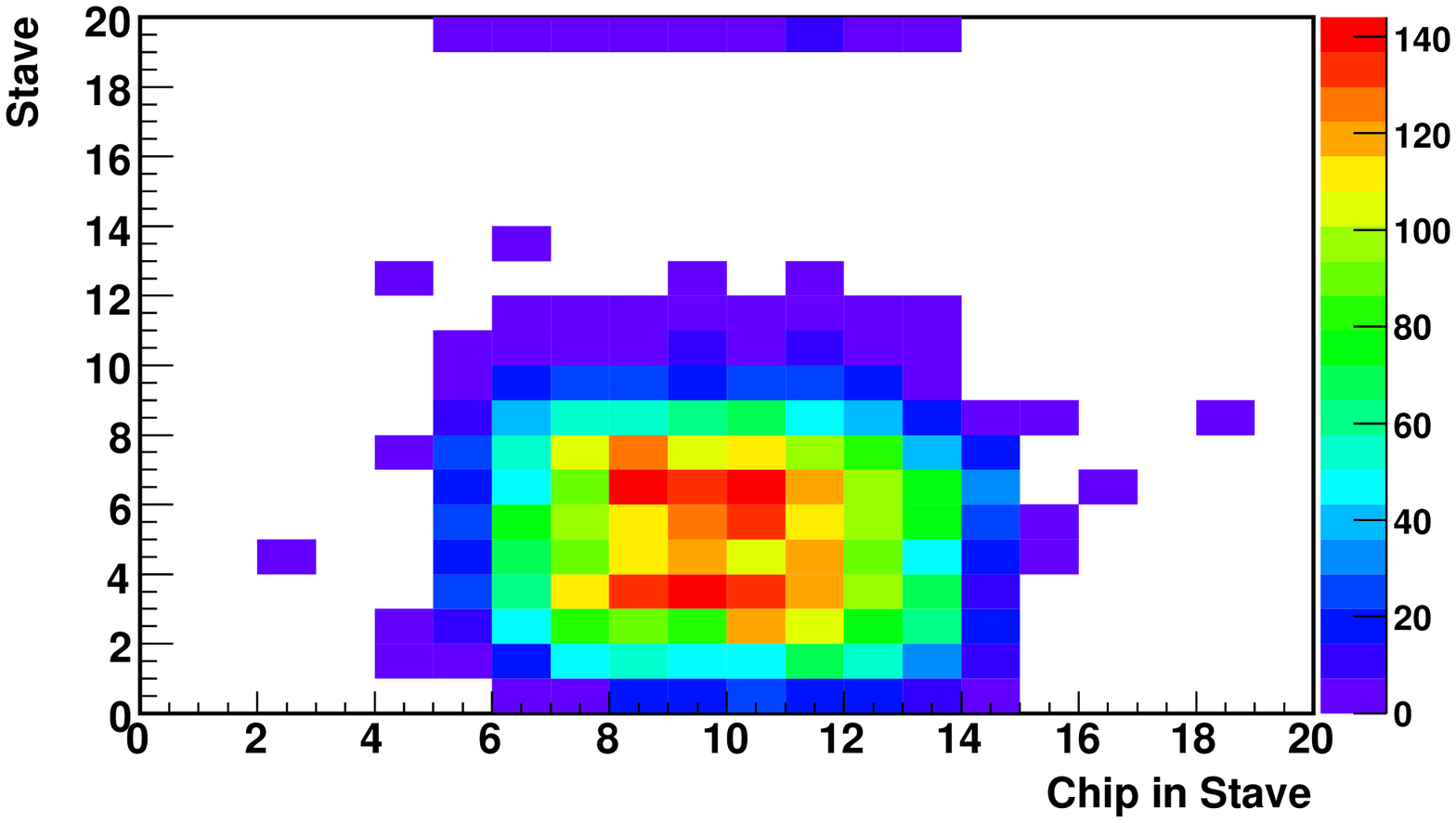,width=6cm}
\psfig{file=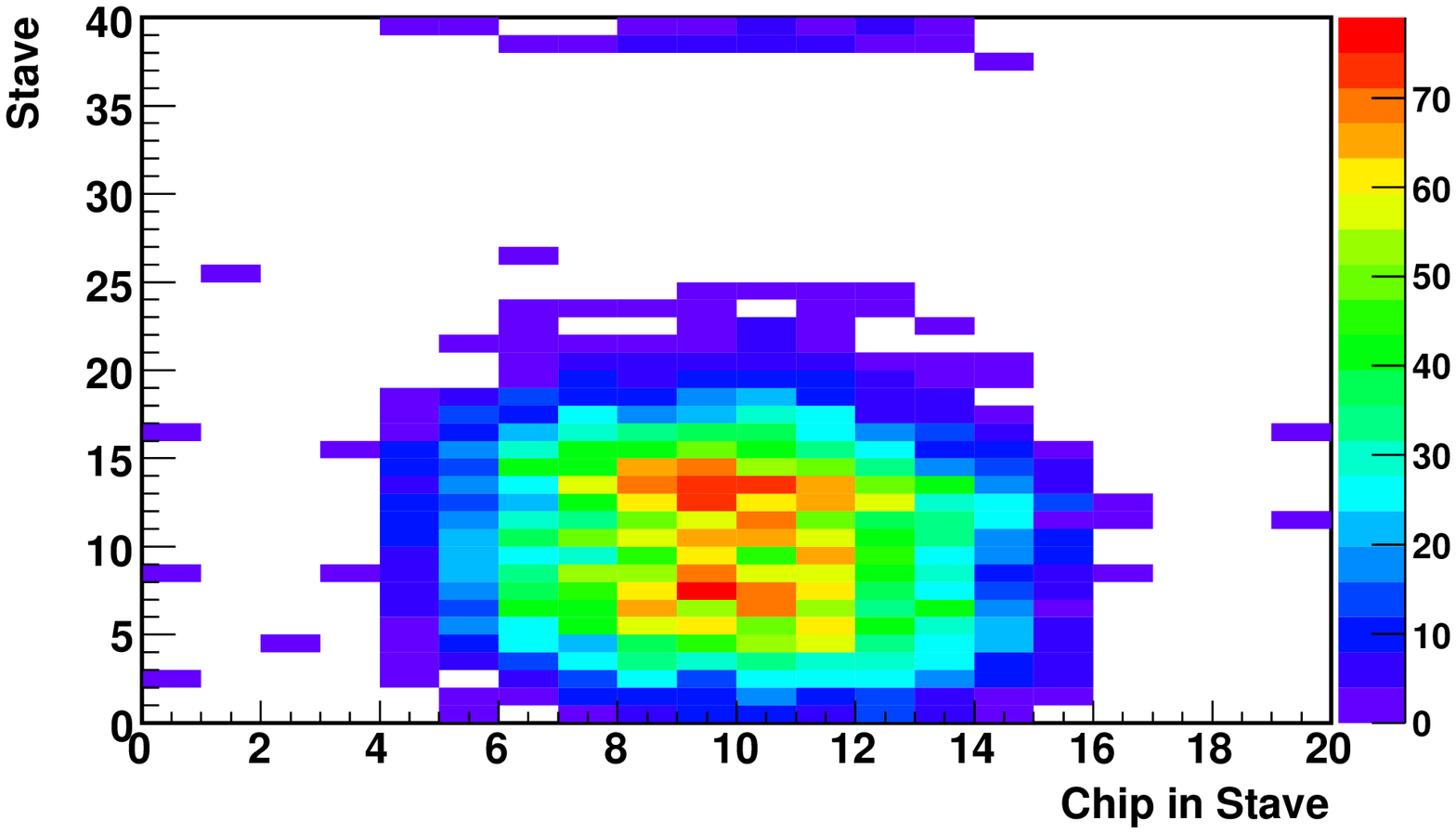,width=6cm}}
\vspace*{2pt}
\caption{Location of the active Fast-OR signals produced by tracks reaching
the HMPID, on the inner (left) and the outer (right) SPD layer.}
\end{figure}

\vspace{-4mm}
A topological condition requiring a minimum number of active
Fast-OR signals in defined fiducial regions onto the inner and
the outer SPD layer 
({\tt R1} and {\tt R2} respectively) can be used. 
As an example the following coincidence condition:
\\

\hspace{10mm} {\tt RT = (nFOinnR1 $>$ nFOminR1) .AND. (nFOoutR2 $>$ nFOminR2)} \\
 
\noindent can be tuned to increase the fraction of interesting events
up to a factor 7.
The bias introduced by such
trigger has to be careful studied
to understand whether selected events may be used for physics.

\section{Conclusion}

The Fast-OR signal available from each of the SPD readout chips 
can be used to contribute to the p-p event selection in ALICE. 
The Pixel Trigger system collects and processes 1200 Fast-OR 
signals, generating an input 
to the CTP. It targets the latency constraint of 800 ns: this
allows to implement Fast-OR based algorithms for the Level 0 trigger.
Both p-p and beam-gas events have been generated and reconstructed 
through the ALICE detector in order to study the performance of
various trigger conditions. Applications to minimum bias, high multiplicity
and special events for the HMPID have been discussed.


\begin{thebibliography}{0}
\bibitem{1} F. Carminati {\em et al.}, 
            J. Phys. G: Nucl. Part. Phys. {\bf 30} (2004), 1517-1763.

\bibitem{2} B. Alessandro {\em et al.},
            J. Phys. G: Nucl. Part. Phys. {\bf 32} (2006), 1295-2040.

\bibitem{3} ALICE ITS Technical Design Report, CERN/LHCC 99-12 (1999).

\bibitem{4} ALICE Trigger Technical Design Report, CERN/LHCC 2003-062 (2004).

\bibitem{5} G. Aglieri Rinella {\em et al.}, 2007 JINST 2 P01007 (2007).

\bibitem{6} ALICE Computing Technical Design Report, CERN/LHCC 2005-018 (2005).

\bibitem{7} T. Sji{\"o}strand {\em et al.}, Computer Physics Commun. {\bf 135} (2001), 238.

\bibitem{8} ALICE Forward Detectors Technical Design Report, CERN/LHCC 2004-025 (2004). 

\bibitem{9} J. Conrad {\em et al.}, ALICE-INT-2005-025 (2005).

\bibitem{10} M. Gyulassy and X.N. Wang, Computer Physics Commun. {\bf 83} (1994), 307.
\end{thebibliography}
\end{document}